\documentclass[reprint,prl,superscriptaddress]{revtex4-1}
\usepackage{graphicx}
 \usepackage{amsmath}
\usepackage{bm}

\begin{document}

\title{Influence of the absorber dimensions on wavefront shaping based on volumetric optoacoustic feedback}

\author{X. Lu\'is De\'an-Ben} \thanks{X.L.D.B and H.S contributed equally to this work}
\affiliation{Institute for Biological and Medical Imaging (IBMI), Helmholtz Zentrum M\"unchen, Neuherberg, Germany}
\author{H\'ector Estrada} \thanks{X.L.D.B and H.S contributed equally to this work}
\affiliation{Institute for Biological and Medical Imaging (IBMI), Helmholtz Zentrum M\"unchen, Neuherberg, Germany}
\author{Ali Ozbek}
\affiliation{Institute for Biological and Medical Imaging (IBMI), Helmholtz Zentrum M\"unchen, Neuherberg, Germany}
\author{Daniel Razansky} \email{Corresponding author: dr@tum.de} 
\affiliation{Institute for Biological and Medical Imaging (IBMI), Helmholtz Zentrum M\"unchen, Neuherberg, Germany}
\affiliation{School of Medicine, Technische Universit\"at München (TUM), Munich, Germany}

\date{\today}

\begin{abstract}

The recently demonstrated control over light distribution through turbid media based on real-time three-dimensional optoacoustic feedback has offered promising prospects to interferometrically focus light within scattering objects. Nevertheless, the focusing capacity of the feedback-based approach is strongly conditioned by the number of effectively resolvable optical modes (speckles). In this letter, we experimentally tested the light intensity enhancement achieved with optoacoustic feedback measurements from different sizes of absorbing microparticles. The importance of the obtained results is discussed in the context of potential signal enhancement at deep locations within a scattering medium where the effective speckle sizes approach the minimum values dictated by optical diffraction.

\end{abstract}


\maketitle 

Light scattering in nanoscale heterogeneities is responsible for the optical opacity of materials, in particular biological tisues \cite{mosk2012}. Even though photons can generally penetrate scattering tissues up to a depth of several centimeters \cite{wang12book}, progressive changes in their propagation direction impede the ability to trace the exact trajectory of light and hence see through this type of samples. The light intensity distribution caused by photon interference is however not entirely random but determined by the type and location of optical scatterers, which results in a characteristic speckle pattern \cite{goodman2007book}. By controlling the incident light wavefront, it becomes possible that photons positively interfere at specific locations, thus making focusing through scattering objects theoretically feasible. Wavefront shaping techniques have recently emerged as a promising tool to image behind strongly scattering samples, with a handful of potential applications foreseen \cite{vellekoop2007,vellekoop2010,mccabe2011spatio,katz2011,katz2014,leonetti2014}. Of particular importance is the possibility to image deep into biological tissues, where scattering limits the penetration of the modern optical microscopy approaches to a depth of a few hundred microns \cite{ntziachristos2010}. In the so-called near-infrared window, diffuse light can penetrate up to several centimeters into highly scattering tissues \cite{wang12book}, thus wavefront-shaping-based techniques can potentially enable high-resolution bio-optical imaging at centimeter-scale depths, ultimately limited by light absorption only.

Wavefront shaping methods are based on the spatial modulation of the phase of the incoming light, which is commonly done by means of a spatial light modulator (SLM) device \cite{mosk2012}. The optimal phase mask at the SLM can be determined from readings of the light intensity at the target point(s). The maximum signal enhancement $\eta$ at a given point (focus) depends on the spatial resolution of the feedback method via the following relation \cite{popoff2011}
\begin{equation} \label{Eq1}
 \eta=\frac{N_{\mathrm{SLM}}}{2N_{\mathrm{modes}}},
\end{equation}
where $N_{\mathrm{modes}}$ is the number of optical modes generating the measured signal and $N_{\mathrm{SLM}}$ is the number of degrees of freedom (pixels) of the SLM. The number of optical modes generally refers to the number of speckles enclosed in the area (or volume) that can be effectively resolved with the technique employed for measuring the light intensity.

The feasibility to focus through scattering objects with wavefront shaping techniques was first demonstrated by using the two-dimensional image acquired with a CCD camera as a means to determine the light energy distribution \cite{vellekoop2007}. Application of this approach to fluorescence microscopy was further showcased \cite{vellekoop2010}. Optical approaches are however not suitable for providing an efficient feedback on the light distribution inside scattering objects as the spatial resolution quickly deteriorates in the diffuse propagation regime \cite{ntziachristos2010}. Thereby, alternative approaches based on a combination of optics and ultrasound have been recently suggested as a means to provide high resolution feedback from a scattering medium. For instance, the frequency (wavelength) of the light beam can be modulated at a given location with focused ultrasound so that the phase of the modulated wavefront can be holographically recorded after propagation through the object. In this case, by means of the phase conjugation, the light beam can be directed to a certain ultrasonically-tagged position \cite{xu2011,si2012,judkewitz2013}. Alternatively, wavefront shaping based on the optoacoustic feedback may offer important advantages to control the scattered light distribution. In this approach, the feedback signals can be in principle collected at a very high rate, ultimately only limited by the ultrasound propagation speed, and hence provide a very fast control mechanism over SLM mask optimization. Acquisition of signals with transducer arrays combined with graphics processing unit (GPU)-based reconstruction procedures further enable real time image rendering in two- and three dimensions \cite{dean2013volumetric}, which can even exceed the currently available refresh rates of standard SLMs \cite{deanben15sr}. Thereby, both single-element transducers and transducer arrays have been suggested for optoacoustic wavefront shaping \cite{kong2011,caravaca2013,chaigne2014controlling,chaigne2014light,dean2015shaping}. Moreover, the high optoacoustic resolution enables a high light intensity enhancement according to Eq.\;\ref{Eq1}, which can be further enhanced by conveniently selecting the detection frequency band \cite{chaigne2014improving} or by using non linear effects \cite{lai2015}. Finally, light absorption essentially occurs in any substance in nature, which makes optoacoustics arguably the most versatile feedback approach in terms if imaging contrast.

Typically, the photothermal energy conversion leading to optoacoustic signal generation takes place in light absorbing molecules. The size and average separation between the molecules is generally much smaller than the speckle dimensions. For example, organic molecules have typical lengths of a few nanometers, while the minimal speckle size is determined by the optical diffraction limit. i.e. $\lambda/2$, where $\lambda$ is the wavelength. For more than one absorber per speckle, $N_{\mathrm{modes}}$ in Eq.\;\ref{Eq1} refers to the number of speckles in the region that can be resolved with the method employed to measure the light intensity. However, larger particles sparsely distributed across the region of interest can also be used to generate the optoacoustic signals, in which case the number of speckles may be generally higher than the number of absorbers. In this case, $N_{\mathrm{modes}}$ in Eq.\;\ref{Eq1} is given by
\begin{equation} \label{Eq2}
 N_{modes}=\sum_{i=1}^{N}N_{sp,i},
\end{equation}
being $N$ the number of particles enclosed in the resolved region and $N_{sp,i}$ the number of optical speckles in each particle. The ratio between the sizes of the speckles and the particles becomes then a crucial factor in determining the potential enhancement achieved, which is also conditioned by the number of particles required to achieve detectable optoacoustic signal levels. Proper selection of the absorber characteristics is therefore of key importance for achieving favorable wavefront shaping results with the optoacoustic feedback approach.

\begin{figure}
	\centering
		\includegraphics[width=0.45\textwidth]{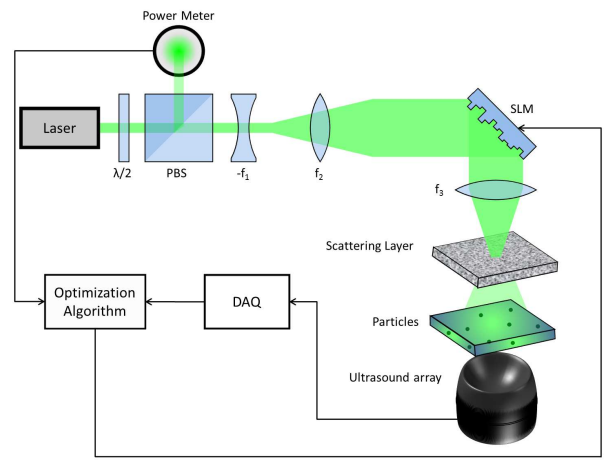}
		\caption{Lay-out of the experimental set-up.}
	\label{Fig1}
\end{figure}

A previously reported volumetric (three-dimensional) optoacoustic wavefront shaping platform \cite{dean2015shaping} was used herein to experimentally investigate the effects of the particle size on the light intensity enhancement. The lay-out of the set-up employed is depicted in Fig. 1. Basically, a phase-only SLM consisting of a liquid crystal on Silicon (LCoS) microdisplay with $8\;\mu\mathrm{m}$ pixel pitch (PLUTO-BB II, Holoeye Photonics AG) was used to control the light wavefront  distribution after passing through a ground glass diffuser (Thorlabs DG10-120). The light enhancement was targeted at polyethylene microparticles with approximate diameters $100\;\mu\mathrm{m}$ (Cospheric BKPMS 90-106) and $200\;\mu\mathrm{m}$ (Cospheric BKPMS 180-210) as well as Carbon microspheres with a diameter around $400\;\mu\mathrm{m}$ (SPI-Supplies) embedded into an agar phantom. The microspheres were excited with the output beam of a frequency-doubled Q-switch Nd:YAG laser (Lab-190-30, Spectral Physics) operating at 15 pulses per second, which was collimated and horizontally polarized before being directed onto the SLM. The generated optoacoustic signals were simultaneously collected with a spherical array of piezoelectric transducers \cite{dean2014functional}, while the three-dimensional optoacoustic images representing the light absorption distribution in the phantom were reconstructed for each laser pulse using a graphics processing unit (GPU) \cite{dean2013volumetric}. 

\begin{figure}
	\centering
		\includegraphics[width=0.45\textwidth]{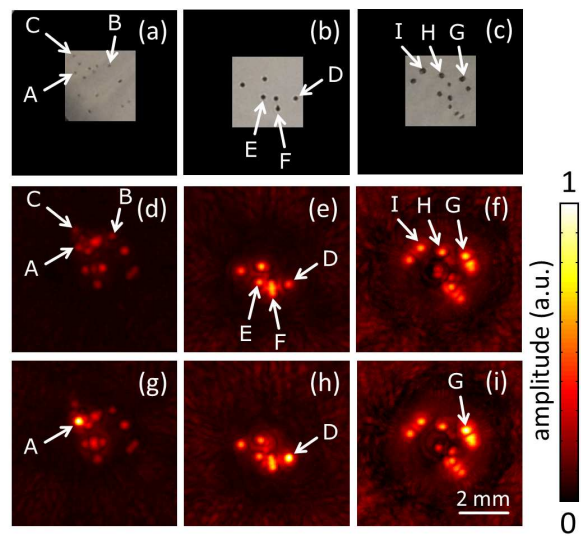}
		\caption{(a-c) Photographs of the agar-embedded microparticles with diameters $100\;\mu\mathrm{m}$, $200\;\mu\mathrm{m}$ and $400\;\mu\mathrm{m}$, respectively; (d-f) Optoacoustic images of the corresponding phantoms obtained with a constant phase value of the SLM pixels (initial iteration). Maximum intensity projections along the depth direction are shown; (g-i) Corresponding optoacoustic images obtained after the phase of the SLM pixels was optimized to deliver maximum light fluence to the particles labelled 'A', 'D' and 'G' respectively.}
	\label{Fig2}
\end{figure}

The reconstructed image was used as a feedback mechanism for a genetic algorithm that optimizes the phase mask at the pixels of the SLM based on the maximization of a defined cost function \cite{dean2015shaping}. In the experiments, a mask of 20x20 phase values was optimized by grouping the SLM pixels accordingly. The speckle diameter was set to approximately $27\;\mu\mathrm{m}$ by controlling the distance between the particles and the diffuser. The cost function was defined as the maximum image value in the volume of interest (VOI) enclosing a certain microsphere, such that the genetic algorithm converges to an SLM mask that focuses light at this particular absorber. The genetic algorithm performs the optimization as described in \cite{caravaca2013}. Specifically, a population of 20 phase masks evolves with a refresh rate of 25\% and a mutation rate that ranges between 10\% and 1.25\% of the pixels of the phase mask. 

Actual photographs of the phantoms are depicted in Figs. 2a-c for microspheres with diameters 100, 200 and 400 $\mu\mathrm{m}$, respectively. The maximum intensity projections (MIP) along the depth direction of the corresponding reconstructed optoacoustic images are subsequently shown in Figs. 2d-f. The MIPs in Figs. 2d-f are obtained by setting a constant phase value in the SLM, which represents the initial iteration of the genetic algorithm. The corresponding MIPs of the images rendered after 2500 iterations of the genetic algorithm are displayed in Figs. 2g-i for cost functions defined to maximize the light intensity on particles labeled A, D and G, respectively. The evolution of the cost function (maximum signal intensity) as a function of the number of iterations in the genetic algorithm for all particles labeled in Fig. 2 is displayed in Figs. 3a-c. Fig. 3d shows the signal enhancement for all the imaged particles as a function of their diameter. The enhancement is estimated as the maximum value of the cost function for the given particle after normalizing it to the average value for the first 20 iterations, which correspond to the initial random population of the genetic algorithm. The small enhancement for the 400 $\mu\mathrm{m}$ particles can be attributed to signal fluctuations approaching the noise levels, so that it was not possible to focus light in those cases. As expected, the enhancement is maximized for smaller particles containing a lower number of speckles.

\begin{figure}
	\centering
		\includegraphics[width=0.45\textwidth]{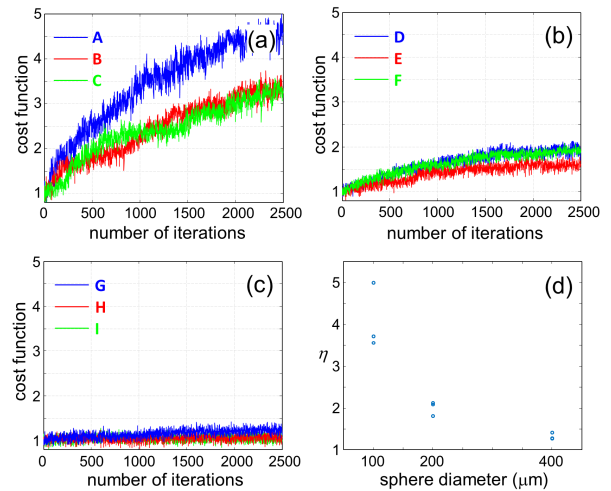}
		\caption{Value of the cost function as a function of the number of iterations in the genetic algorithms for the particles with diameter $100\;\mu\mathrm{m}$ (a), $200\;\mu\mathrm{m}$ (b) and $400\;\mu\mathrm{m}$ (c). The light intensity enhancements achieved as a function of the particle diameter are displayed in (d).}
	\label{Fig3}
\end{figure}

The results showcased indicate the importance of the absorber dimensions used to provide feedback in optoacoustics-based wavefront shaping, which ultimately determines the focusing (light enhancement) capacity of this approach. The maximum signal enhancement achieved is of particular relevance when eventually aiming at focusing light inside scattering samples, e.g. biological tissues, where the speckles at depths beyond the transport mean free path are expected to have characteristic lengths approaching $\lambda$/2. In order to generate efficient feedback from absorbers in the sub-micron size range, the optoacoustic imaging system would need to have an efficient detection bandwidth in the hundreds of MHz range. However, the image artifacts and resolution degradation associated with acoustic heterogeneities and attenuation of the high frequency components in the imaged medium are expected to impose hard limitations on the achievable spatial resolution when using the optoacoustic imaging feedback. In biomedical optoacoustic imaging, the practically achievable resolution is generally in the range of 1/200 of the imaging depth \cite{wang2012photoacoustic}. Thus, the realistically achievable spatial resolution at a depth of 5mm is expected to be in the range of $25\;\mu\mathrm{m}$. As a result, approximately 250000 speckles will be accommodated in each resolution-limited voxel, for which no significant light intensity enhancement is feasible. Higher efficiency of the light enhancement can be potentially accomplished when properly selecting the absorbing particles and their concentration so that a relatively low number of light speckles is enclosed within the resolved volume while simultaneously attaining sufficient detection sensitivity. Yet, focusing light within biological tissues still remains a challenging topic due to additional factors, such as the millisecond-level speckle decorrelation times.

In conclusion, the size of light absorbing particles is designated to play an essential role in volumetric wavefront shaping using optoacoustic feedback. The experimental analysis performed in this work may help to provide general guidelines for optimizing the particle size for achieving efficient light enhancement at locations exhibiting small speckle size, e.g. beyond the transport mean free path depth within turbid tissues.

This project was supported in part by the European Research Council through the grant agreement ERC-2010-StG-260991.

\end{document}